\documentclass[sigconf]{acmart}

\usepackage{balance}
\usepackage{graphicx}
\usepackage{diagbox}
\usepackage{subfigure}
\usepackage{amsmath}
\usepackage{adjustbox}
\usepackage{xcolor}
\usepackage{stmaryrd}
\usepackage{hyperref}
\usepackage{enumitem}
\usepackage{wrapfig}
\usepackage{mathtools}
\usepackage[linesnumbered,algoruled,boxed,lined]{algorithm2e}
\usepackage{algpseudocode}
\DeclareUnicodeCharacter{2212}{-}
\DeclareUnicodeCharacter{FF1A}{-}
\usepackage[utf8]{inputenc} 
\usepackage[T1]{fontenc}    
\usepackage{hyperref}       
\usepackage{url}            
\usepackage{booktabs}       
\usepackage{amsfonts}       
\usepackage{nicefrac}       
\usepackage{microtype}      
\usepackage{xspace}

\usepackage{appendix}
\usepackage{amsthm}
\usepackage{mathtools, nccmath}

\usepackage{multirow}
\usepackage{float}
\usepackage{caption}
\usepackage{ctable}
\theoremstyle{definition}

\AtBeginDocument{%
  \providecommand\BibTeX{{%
    \normalfont B\kern-0.5em{\scshape i\kern-0.25em b}\kern-0.8em\TeX}}}
    
\renewcommand\footnotetextcopyrightpermission[1]{} 
\acmConference[]{}{}{}

\author{Junyao Zhang}
\affiliation{%
  \institution{University of Southern California}
  \country{Los Angeles, CA}
  }
\email{junyaozh@usc.edu}

\author{Paul Bogdan}
\affiliation{%
  \institution{University of Southern California}
  \country{Los Angeles, CA}
  }
\email{pbogdan@usc.edu}

\author{Shahin Nazarian}
\affiliation{%
  \institution{University of Southern California}
  \country{Los Angeles, CA}
  }
\email{shahin.nazarian@usc.edu}

\begin{document}

\title{C-SAR: SAT Attack Resistant Logic Locking for RSFQ Circuits}
\begin{abstract}
Since the development of semiconductor technologies, exascale computing and its associated applications have required increasing degrees of efficiency. Semiconductor-transistor-based circuits (STbCs) have struggled in increasing the GHz frequency. Emerging as an alternative to STbC, the superconducting electrons (SCE) technology promises higher-speed clock frequencies at ultra-low power consumption. The rapid single flux quantum (RSFQ) circuits have a theoretical potential for three orders of magnitude reduction in power while operating at clock frequencies higher than 100 GHz. Although the security in semiconductor technology has been extensively researched and developed, the security design in the superconducting field requires field demands attention. In this paper, C-SAR is presented that aims to protect the superconducting circuit electronics from Boolean satisfiability (SAT) based attacks. The SAT attack is an attack that can break all the existing combinational logic locking techniques. C-SAR can immunize against SAT attacks by increasing the key search space and prolonging the clock cycles of attack inputs. Even in the worst case of C-SAR, in face of S-SAT a specially designed SAT attack, C-SAR can also soar the attack cost exponentially with key bits first, then linearly with the length of camouflaged DFF array. We have shown in this work that the cost of C-SAR is manageable as it only linearly increases as a function of key bits. 
\end{abstract}


\maketitle
\section{introduction}
As a strong competitor to the semiconductor circuits in the field of super electronic products, the SCE has shown great potential in ultra-high-speed clock frequency and ultra-low power consumption \cite{RSFQ}. The fabrication process of Nb/Al-AlO$_X$/Nb Josephson junction based on 200nm production-level tools has been developed \cite{fabrication}. As the most popular member of the SCE family, RSFQ has a similar design flow as traditional CMOS. This similarity not only enables RSFQ to become a compelling technology for SCE and an expected substitute for CMOS, but also exposes the RSFQ circuit to malicious attacks \cite{reverse RSFQ}.

In the semiconductor area, the growing complexity of integrated circuits (ICs) and the soaring cost of building or maintaining semiconductor foundries have driven the globalization of IC design and manufacturing\cite{DSB}. Consequently, malicious agents can use this global supply chain to illegally access IPs or physical ICs, posing a threat to the entire supply chain market and causing the industry to lose \$$4$ billion annually \cite{SEMI}. These illegal activities include reverse engineering (RE) of IC netlists, inserting malicious parts into circuits, overbuilding ICs, \textit{etc} \cite{Trustworthy,Camouflaging}. Logical locking techniques can thwart the above unauthorized use of IPs from untrusted agents by embedding secret keys \cite{Third_shift}. 

Recently, a Boolean satisfiability (SAT) attack was demonstrated to exploit the specific weakness of the logic locking technology \cite{evaluating encryption}. The objective of a malicious agent is to extract the secret key. And they can access both locked netlists with the key gates, and the functional IC embedding correct keys. SAT-based attacks iterate over each input pattern applied to the IC and use SAT solvers to rule out incorrect key sets by calculating the outputs. The attack finishes when all the incorrect key sets are eliminated \cite{evaluating encryption}. SAT-based attacks can crack the keys of most existing combinational logic locking technologies in a short time \cite{evaluating encryption}.

\begin{figure}[t]
    \centering
        \includegraphics[width=0.42\textwidth]{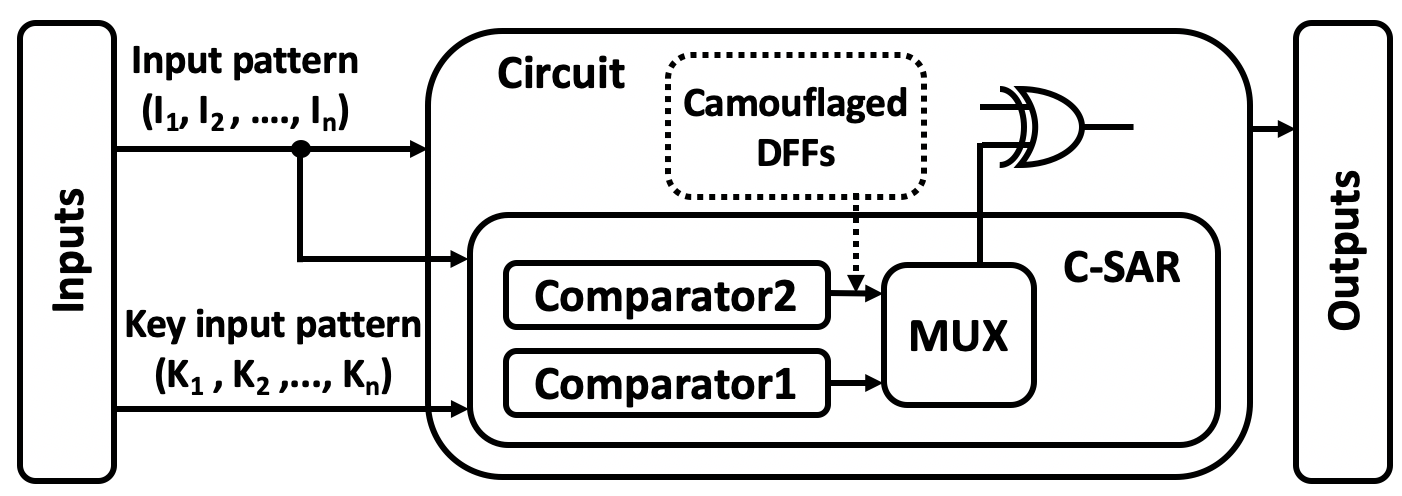}
        \caption{Overview of C-SAR logic locking, C-SAR extends the required clock cycles and reduces the number of key values filtered in each attack iteration to immunize against SAT-based attacks. The camouflaged DFF array intertwines with C-SAR to resist removal attacks}
        \label{fig:overview}
\end{figure}

Since the principles of fabricating RSFQ are similar to CMOS \cite{fabrication}, the above RE methodology is possible to apply to the RSFQ logic. However, the threat to RSFQ circuits will be more significant due to the relatively large size devices and a small number of devices per IC, as compared to CMOS \cite{reverse RSFQ}. Although the security of circuits in the CMOS field has been extensively studied, the safety issues in the SCE field remain further explorations and developments. Promoting the security of SCE circuits is the focus of this research. To summarize, the main contributions of this paper are as follows:
\begin{itemize}[leftmargin=*]
    \item We propose a logic locking technology called C-SAR (see Figure \ref{fig:overview}) to countermeasure SAT attacks in superconducting circuits.
    \item We couple the camouflaged DFF array with C-SAR to obfuscate the input temporal distribution to hide its functional block. This can resist removal attacks which intend to identify and separate the C-SAR block from the original netlist. 
    \item We validate C-SAR in ISCAS’85 and 74X-series benchmark suites. Rigorous analysis and experimental results show that it can immunize against SAT attacks. Our approach has high effectiveness in thwarting S-SAT attack (explained in section \ref{sec:secure analysis}), a specially designed SAT attack. The attack effort increases exponentially with the number of key bits, then grows linearly with the number of camouflaged DFFs.
\end{itemize}

\section{background}
\subsection{Rapid single flux quantum logic (RSFQ)}\label{sec:RSFQ}
In RSFQ, overdamped Josephson junctions using resistive current bias are connected such that the binary information is present in a picosecond duration of voltage pulses $V(t)$, instead of voltage levels in semiconductor technologies. Since the voltage pulse is equivalent to a quantum of flux $\phi_o= 2.07{\times}10^{−15} V{\cdot}s$, it is also called single flux quantum pulse (SFQ pulses) \cite{RSFQ}.
These unique properties alter the conventional understanding of the representation of bits. The basic convention is that the arrival of an SFQ pulse during the current clock period represents a logic “$1$” whereas the absence of any pulse during this period is understood as logic “$0$”. The convention only requires that a pulse arrives sometime during the clock rather than the sequence of the input pulses and exact time coincidence of the input pulses in the period \cite{RSFQ}. The output pulses cause a reaction to input pulses only if the clock pulse arrives. Therefore, it can be considered the same as a CMOS gate with an edge-triggered flip-flop at the output. Moreover, the RSFQ gates can be divided into two parts according to whether it is controlled by a clock. For example, a Josephson transmission line (JTL), splitter and a confluence buffer do not require a clock. In addition, the propagation delay is the delay of the output with respect to the input. For example, logic gates such as NOT, OR, AND, and XOR are clocked. Hence, the propagation delay is measured as the time elapsed after the clock pulse arrives \cite{SFQ_map}.

\subsection{Logic locking}
\begin{figure}[h]
    \centering
    \subfigure[]{
    \includegraphics[width=0.45\linewidth]{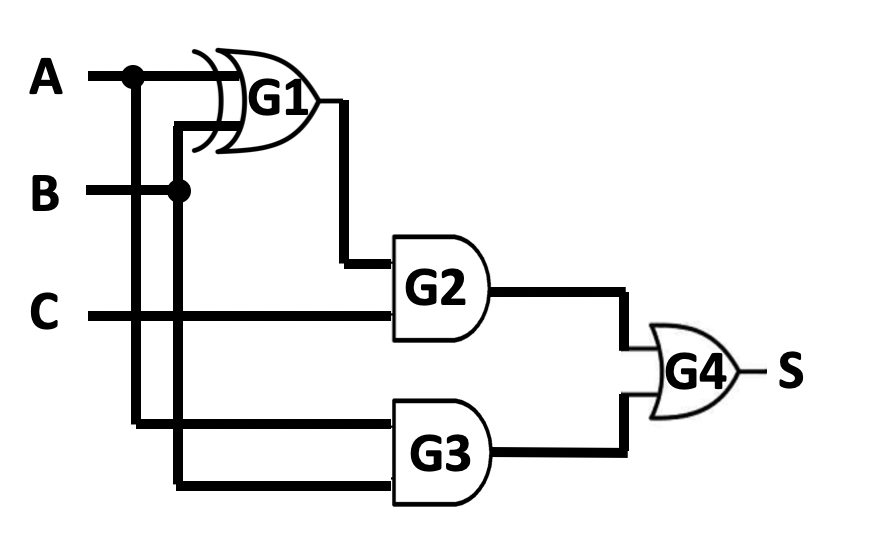}
    }
    \subfigure[]{
    \includegraphics[width=0.45\linewidth]{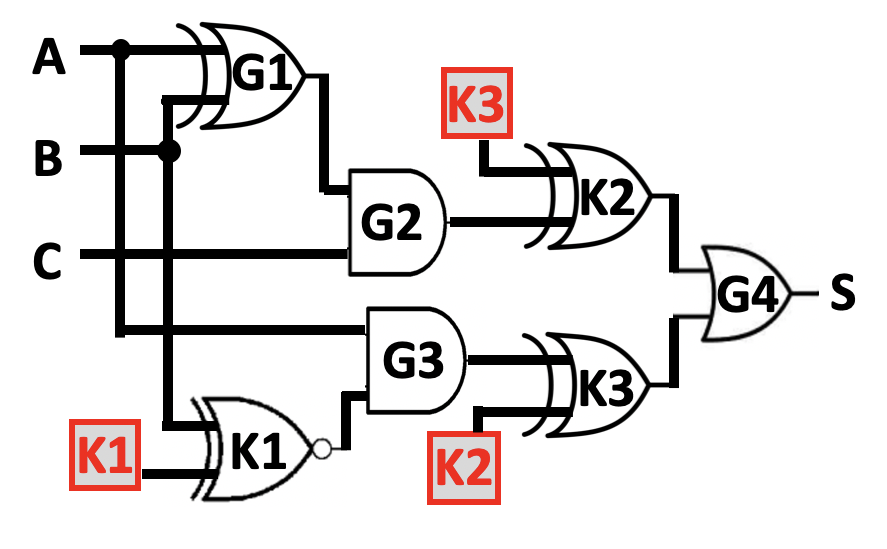}
    }
    \caption{\textbf{a.} Circuit example with 3 inputs; \textbf{b.} XOR/XNOR key gates logic locking circuit \cite{EPIC}. Correct key: $K3K2K1 =001$}
    \label{fig:circuit set}
\end{figure}

Figure \ref{fig:circuit set}.a and Figure \ref{fig:circuit set}.b show a netlist and the locked version (by adding three XNOR/XOR key gates \cite{EPIC}), respectively. In the original design, there are only the primary inputs that are driven by wires. In the locked version, the key inputs are the newly added inputs, which are driven by key bits stored in the tamper-proof memories \cite{evaluating encryption}. A locked netlist cannot generate the correct output unless activated with the correct keys \cite{evaluating encryption}. XOR and XNOR can replace each other to improve the complexity of the correct key set.

\begin{figure}[t]
    \centering
        \includegraphics[width=0.45\textwidth]{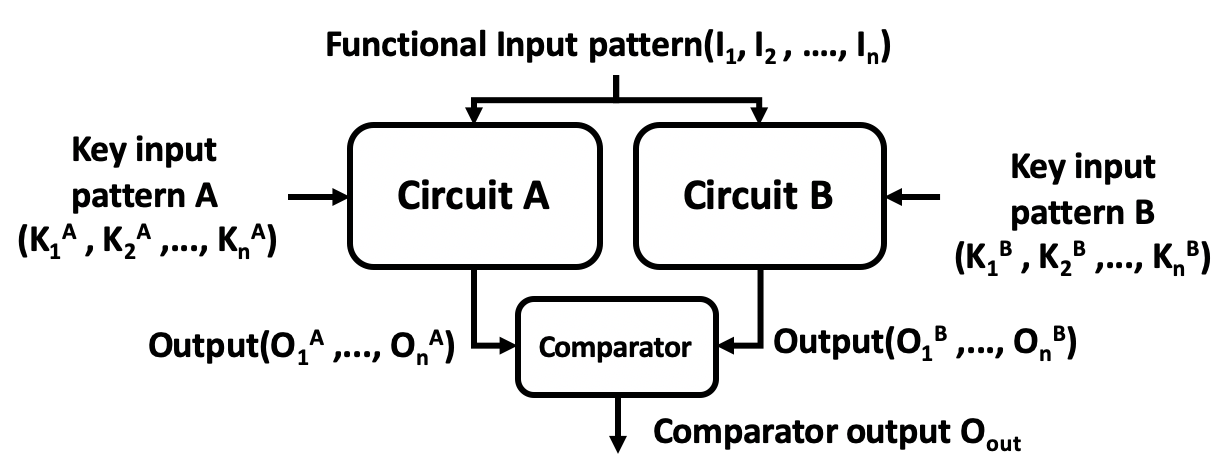}
        \caption{SAT attack example circuit}
        \label{fig:SAT attack circuit}
\end{figure}

\subsection{SAT attack}\label{sec:SAT}
The SAT-based attack iterates to eliminate the incorrect key values by distinguishing input patterns \cite{evaluating encryption}. Figure \ref{fig:SAT attack circuit} illustrates a circuit example of the SAT algorithm \cite{SARLock}. Two groups of identical locked circuit copies share the same input pattern $[I_1\dots I_n]$ with different key sets $[K_1^A \dots K_n^A]$ and $[K_1^B \dots K_n^B]$. The attacker applies circuit outputs to the comparator, which generates $O_{out} = 1$ if $[O_1^A \dots O_n^A]$, $[O_1^B\dots O_n^B]$ are different, or $0$ for identical values. Then, the SAT solver applies this input pattern $[I_1\dots I_n]$ to the functional IC to obtain the correct output $[O_1^C\dots O_n^C]$ when outputs of locked circuit copies are different. The solver can rule out the key sets with incorrect output. Thus, it is possible to rule out all the incorrect key sets with one input pattern. If the correct key set still hides, the attacker adds more distinguishing input patterns to the circuit until no further input pattern is left. It indicates that the attack is successful if all the incorrect key sets are ruled out \cite{evaluating encryption}. 

\begin{table}[t]
    \centering
    \caption{Analysis of the SAT-based attack against logic locking: 
    Column $ABC$ represents different input patterns; $k_0$-$k_7$ show the locked circuit’s output for different key sets; $S$ denotes the correct output. The correct key set is $k_1=001$}
    \label{tab:truthtable}
    \begin{tabular}{|c|c|c|c|c|c|c|c|c|c|}
    \hline
    \hline
    ABC & S & $k_0$ & $k_1$ & $k_2$ & $k_3$ & $k_4$ & $k_5$ & $k_6$ & $k_7$ \\\hline
    000 & 0 & 0 & 0 & 1 & 1 & 1 & 1 & 1 & 1\\\hline
    001 & 0 & 0 & 0 & 1 & 1 & 1 & 1 & 1 & 1\\\hline
    010 & 0 & 0 & 0 & 1 & 1 & 1 & 1 & 1 & 1\\\hline
    011 & 1 & 1 & 1 & 1 & 1 & 0 & 0 & 1 & 1\\\hline
    100 & 0 & 1 & 0 & 0 & 1 & 1 & 1 & 1 & 1\\\hline
    101 & 1 & 1 & 1 & 1 & 1 & 1 & 0 & 0 & 1\\\hline
    110 & 1 & 0 & 1 & 1 & 0 & 1 & 1 & 1 & 1\\\hline
    111 & 1 & 0 & 1 & 1 & 0 & 1 & 1 & 1 & 1\\\hline
    \end{tabular}
\end{table}

\begin{figure}[b]
    \centering
        \includegraphics[width=0.35\textwidth]{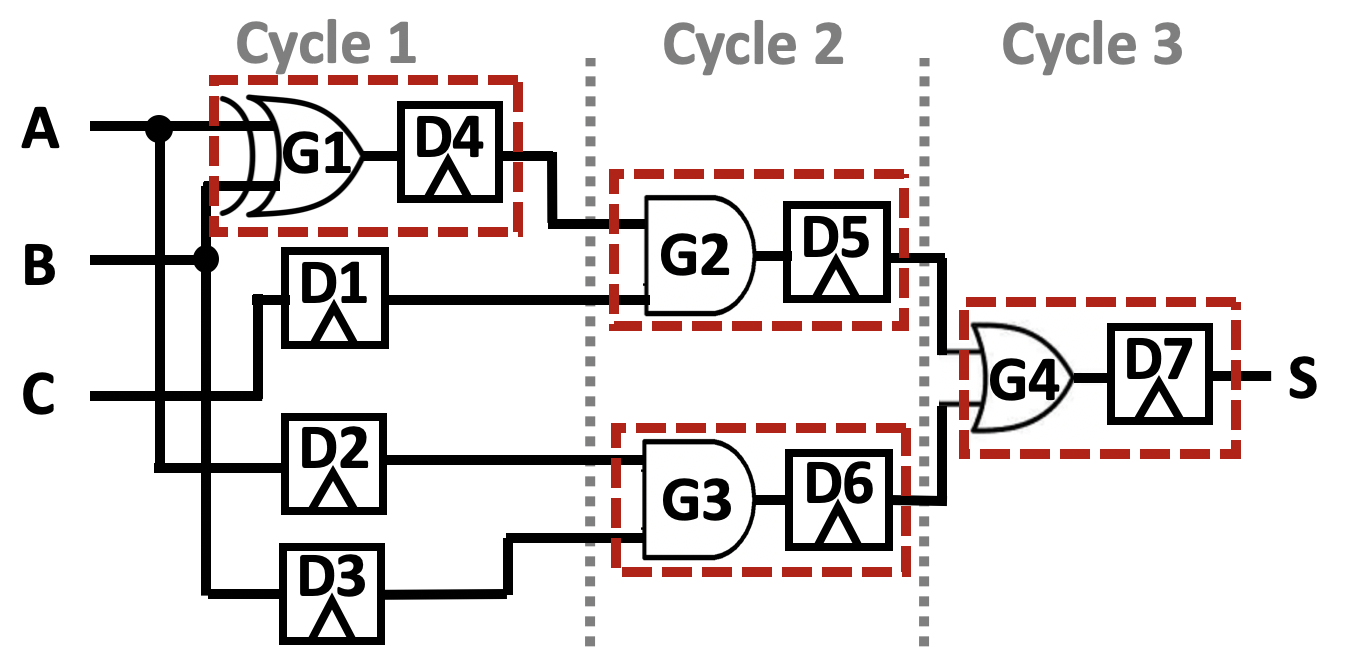}
        \caption{Equivalent RSFQ circuit of example in Figure \ref{fig:circuit set}.a}
        \label{fig:equivalent circuit}
\end{figure}

The locked circuit in Figure \ref{fig:circuit set}.b is used to analyze the process of SAT-based attacks. Table \ref{tab:truthtable} represents the corresponding output values for different key sets and input patterns. $k_0$ to $k_7$ represent the $8$ possible different key sets as the key bit is $3$ ($k_4$ means \{K3K2K1=100\}). Assuming the input pattern $101$ is applied in the first iteration, the key set $k_5$ and $k_6$ are ruled out due to their outputs are different from the correct result "1" in $S$ shown in Table \ref{tab:truthtable} row $101$. In the next iteration, $111$ is applied, $k_3$ and $k_0$ are pruned by a similar comparison as the previous iteration. In the third iteration, $000$ can rule out $k_2$, $k_4$, $k_7$. Therefore, $7$ incorrect key sets are all eliminated out in $3$ iterations. Thus, this SAT-based approach extracts the correct key set $k_1$.

\section{C-SAR logic locking technique}
Section \ref{sec:RSFQ} briefly explains the functioning of RSFQ gates, except that the splitters and JTL require a clock signal to transfer the stored quantum flux to their outputs. It means that the RSFQ circuits must be completely gate-level pipelined. Figure \ref{fig:circuit set}.a is mapped to an RSFQ circuit (Figure \ref{fig:equivalent circuit}) as an example. $D1$, $D2$, and $D3$ are added to equalize the logic depth as Path-balancing D-flip flops \cite{SFQ_map}. The equivalent circuit requires 3 clock cycles to produce its output rather than 1 clock cycle.

\subsection{Configuration of RSAT block}\label{sec: RSAT}
\begin{figure}[t]
    \centering
        \includegraphics[width=0.45\textwidth]{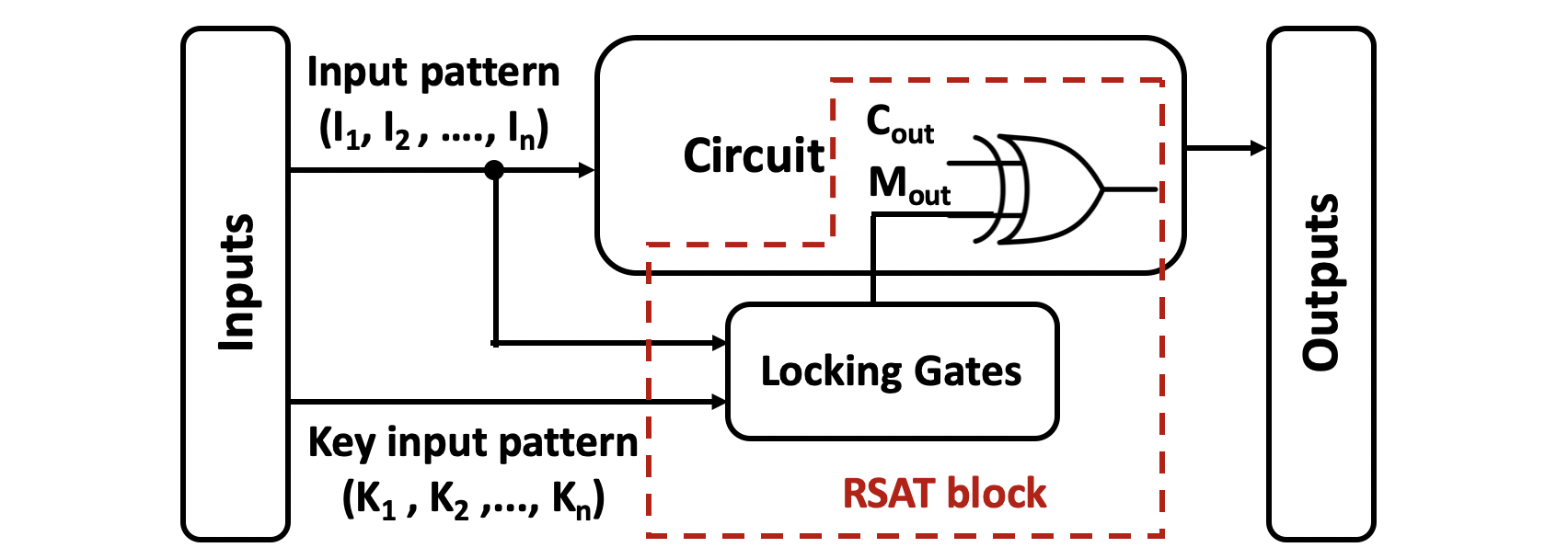}
        \caption{Integrating RSAT resist block into a circuit}
        \label{fig:Integrating RSAT}
\end{figure}

The success of SAT-based attack against the existing logic locking techniques is the fact that these technologies fail to take into account the discriminating ability of individual input patterns \cite{SARLock}. As in the previous example in Table \ref{tab:truthtable}, when the input pattern is $100$, all incorrect key sets except $k_2$ can be excluded. This is a relatively effective scenario for SAT-based attacks to prune multiple incorrect key sets through a single input pattern. Consequently, we propose a logic locking technology named RSAT, which can operate the circuit output to keep at most one wrong key set having the distinguishing output in each SAT attack attempt. Based on the above reasoning, a SAT-based attack resistance circuit can be built as Figure \ref{fig:Integrating RSAT}.

\begin{figure}[b]
    \centering
        \includegraphics[width=0.45\textwidth]{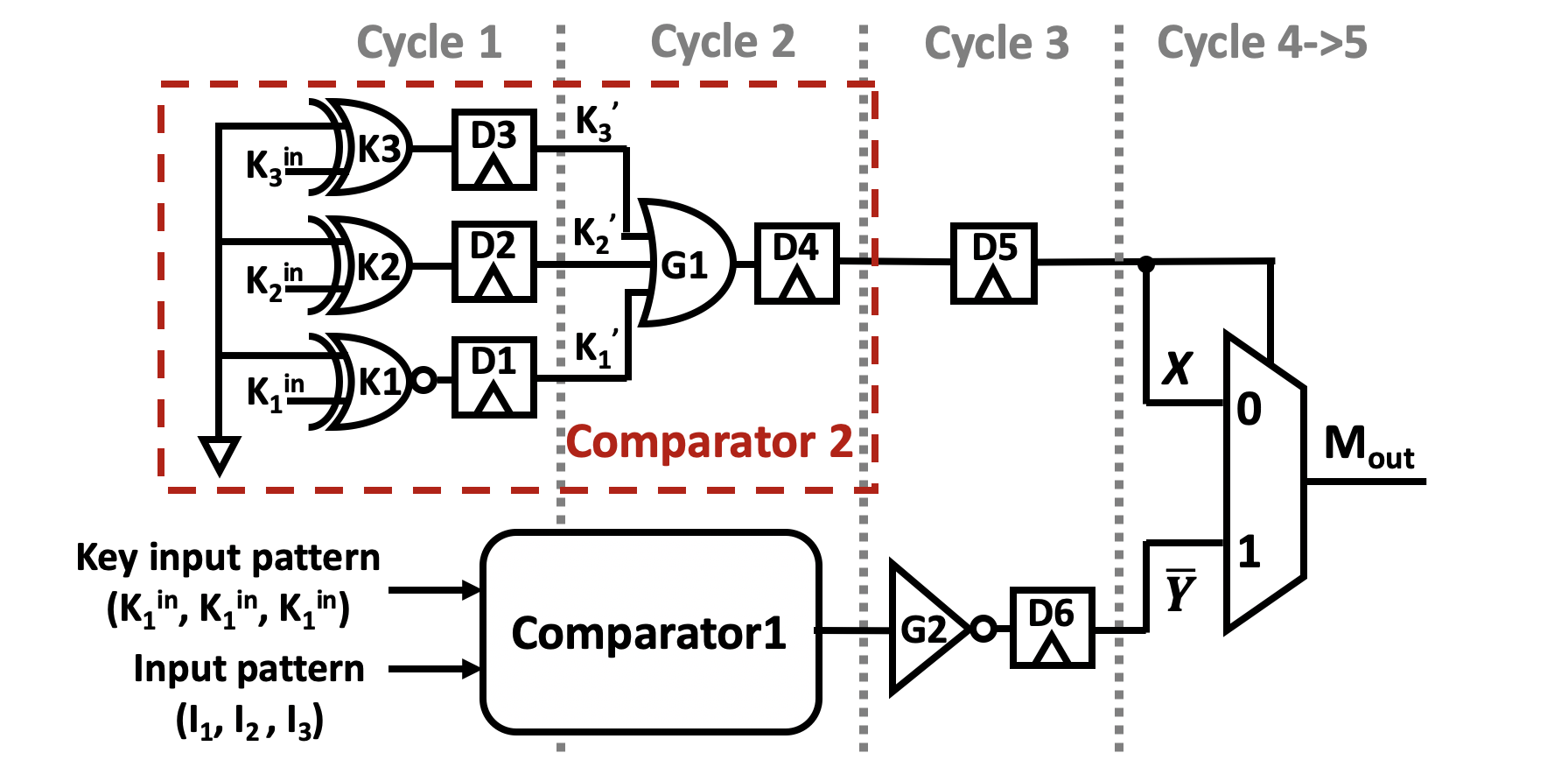}
        \caption{RSFQ logic based RSAT circuit, correct key is $K_1^{c}K_2^{c}K_3^{c}=001$}
        \label{fig:RSAT}
\end{figure}

An example of locking gates in the RSAT block is implemented in RSFQ logic with the correct key set $[0, 0, 1]$, shown in Figure \ref{fig:RSAT}. The ground value and key set inputs are XOR-ed or XNOR-ed to form $K_1^{'}$, $K_2^{'}$ and $K_3^{'}$. As the sum of signals from $K_1^{'}$ to $K_3^{'}$, $X$ performs both the control value of the MUX, and the input value of the MUX when the control value is $0$. It will take 2 clock cycles to generate $X$ after a key set is driven into the circuit. Another input of the MUX is $\Bar{Y}$, the reverse value of the comparator 1 output. $K1$, $K2$ and $K3$ can be replaced with XNOR/XOR gates to increase the degree of complexity and confusion. This makes it harder for an attacker to know whether an inverter is part of original netlist or is added for logic locking. When the input key set $[K_1^{in}, K_2^{in}, K_3^{in}]$ is identical to correct key set $[K_1^{c}, K_2^{c}, K_3^{c}]$ ($[0, 0, 1]$ in Figure \ref{fig:RSAT}), the MUX gate selects $X$, otherwise $\Bar{Y}$ is used. Then, the MUX takes another 2 clock cycles to output $M_{out}$. The logic of $M_{out}$ is summarized as:
\begin{equation}
M_{out} = \Bar{Y}\cdot(K_n^{c}\otimes K_n^{in}+, \dots, K_1^{c}\otimes K_1^{in})
\end{equation}

The comparator 1 outputs 1 for different inputs. This block takes 2 clocks to output $Y$, regardless of whether it uses XNOR-AND or XOR-OR structure. The inverse value of $Y$ requires an extra clock. In summary, from input pattern and key input pattern to $\Bar{Y}$ needs 3 clocks. DFF $D5$ is added at $X$ to balance the needed clock cycles.

In this design, the primary output of the circuit $C_{out}$ can only be flipped when the specific combinations of input pattern and key set are inputted. The RSAT block can protect the circuit by reducing the number of exposed incorrect key sets to 1 in each SAT attack attempt. The SAT attack needs $2^{N_{key}}$ clock cycles to prune all the incorrect key sets, when the key bit is $N_{key}$.

\subsection{C-SAR: camouflaged SAT attack resist} \label{sec:C-RSAT}
\subsubsection{Intertwine the RSAT with other defenses}
With the increasing scale and complexity of the circuits, the logic depth of the circuits also becomes larger. It is necessary to force $M_{out}$ taking the same clocks to reach the XOR gate as the execution time of the original circuit. This can guarantee the correct functionality of RSAT block. Therefore, the RSAT block needs DFF arrays to equalize the logic depth of $M_{out}$ to $C_{out}$. However, the attacker can extract the unique path-balancing flip-flop mapping of the RSAT block by examining the netlist layout \cite{reverse RSFQ}. Path-balancing flip-flop mapping can decipher the temporal distribution of the inputs, and then determine the circuit functionality \cite{reverse RSFQ}. Consequently, the attacker can separate the RSAT block from the original circuit with this principle. To defend against such removal attacks, the RSAT block in section \ref{sec: RSAT} needs to be coupled with other defense mechanisms. 
\begin{figure}[t]
    \centering
    \subfigure[]{
    \includegraphics[width=0.45\linewidth]{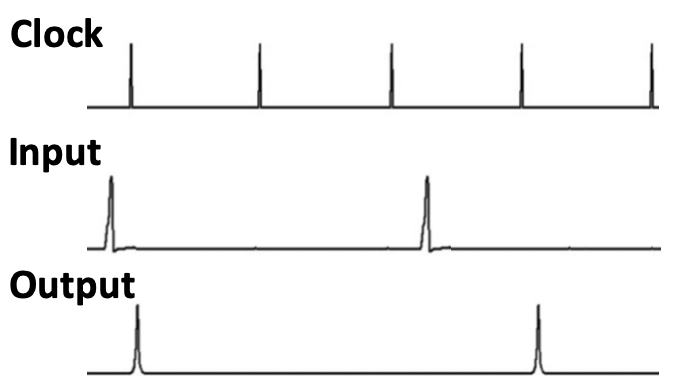}
    }
    \subfigure[]{
    \includegraphics[width=0.45\linewidth]{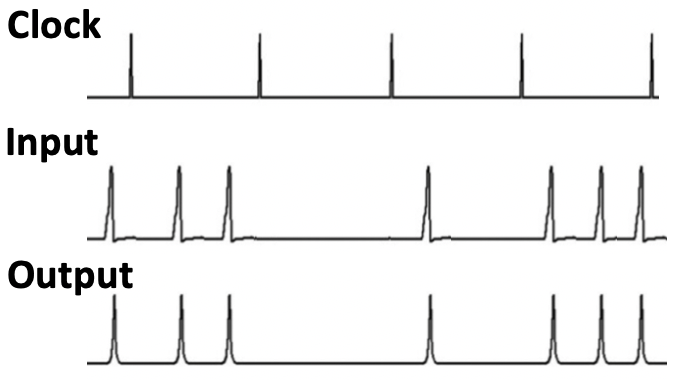}
    }
    \caption{\textbf{a.} Standard DFF in RSFQ logic. \textbf{b.} Camouflaged DFF operating as a JTL \cite{reverse RSFQ}}
    \label{fig:camouflaged DFF simulation}
\end{figure}

\begin{figure*}[ht]
    \centering
        \includegraphics[width=0.70\textwidth]{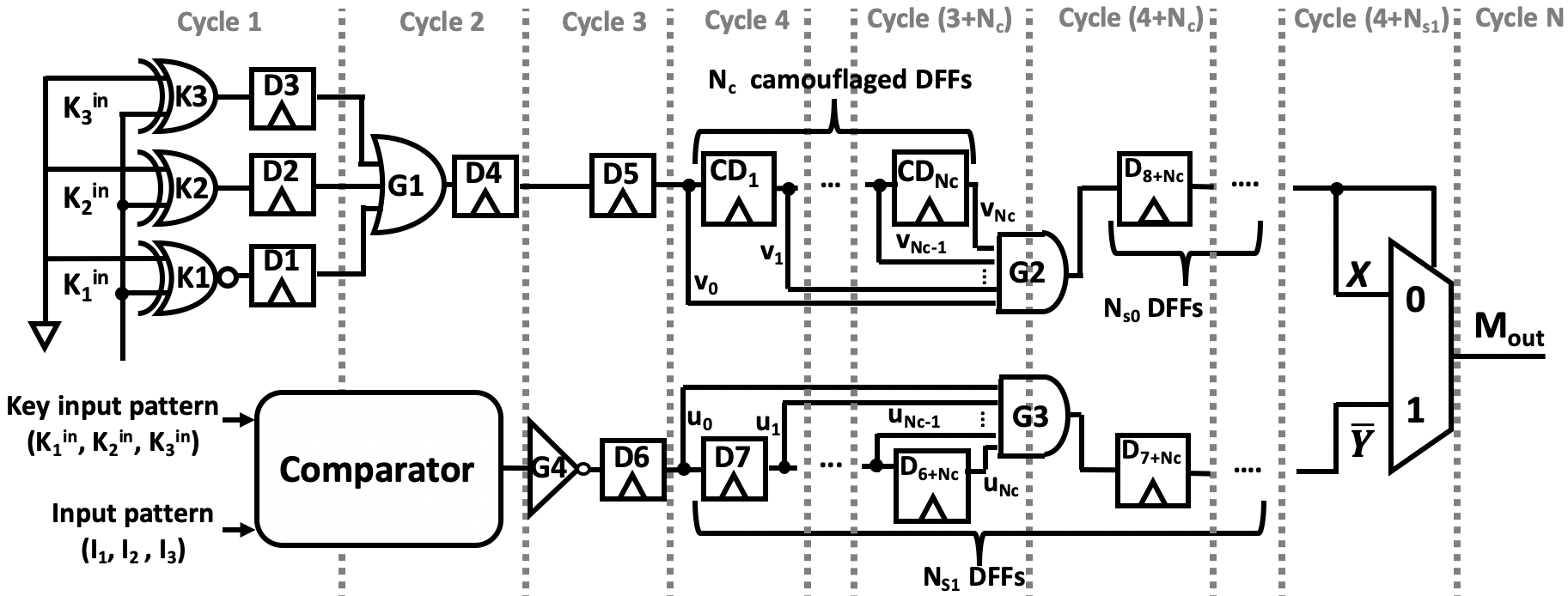}
        \caption{Configuration of C-SAR technique, CD is camouflaged DFF, D is standard DFF, correct key is $K_1^{c}K_2^{c}K_3^{c}=001$}
        \label{fig:C-RSAT}
\end{figure*}

\subsubsection{Camouflaged RSFQ D-Flip Flop}\label{sec:C_DFF}
In the existing camouflaged RSFQ cells, the camouflaged RSFQ DFF can be used as a JTL or a standard DFF. The camouflaged DFF can be converted from the standard DFF by adjusting the thickness of the insulating layer and replacing the Josephson junctions with dummy Josephson junctions. It has comparable energy dissipation and area overhead with a standard DFF but functions as a JTL \cite{reverse RSFQ}. These features can complicate the deciphering process of the attacker. The simulation of the camouflaged DFF is shown in Figure \ref{fig:camouflaged DFF simulation}.b. Figure \ref{fig:camouflaged DFF simulation}.a is the standard DFF. As explained in section \ref{sec:RSFQ}, the camouflaged DFF introduces zero clock cycle delay as JTL. It passes the input pulses regardless of the clock. A standard DFF introduces a single clock cycle delay. The attacker cannot deduce the function of the camouflaged netlist due to an exponential number of possible input combinations introduced by the camouflaged DFF array. Thus, the camouflaged DFF can hide the unique temporal distribution of the inputs of the circuit to protect the RSAT block from being removed. Moreover, the camouflaged DFFs are added in different pipeline levels of a circuit. It creates a different clock delay in each circuit path, which causes a temporal obfuscation at outputs.

\subsubsection{Configuration of C-SAR}
We integrate our RSAT logic locking with camouflaged DFF array to resist the SAT attacks and removal attacks. It is named C-SAR for short and depicted in Figure \ref{fig:C-RSAT}. The clock cycles of both MUX inputs are balanced by a standard DFF $D5$. Both paths will take 6 clock cycles without DFF arrays:
\begin{equation} \label{eq: constant}
T=
\begin{cases}
\text{ $X = 0, 2(MUX) + 1(XOR) + 1(OR) + 1(DFF) + 1(AND)$}\\ 
\text{ $X = 1, 2(MUX) + 2(Comparator) + 1(INV)+ 1(AND)$}
\end{cases}
\end{equation}
where $N_c$ camouflaged DFFs and $N_{s0}$ standard DFFs are inserted at $X$ as path-balancing DFFs. In the other path $\Bar{Y}$, $N_{s1}$ standard DFFs are placed, when the original circuit requires $N$ clock cycles to output its result. The following equations (\ref{eq: N}) and (\ref{eq: Ns1}) indicate the quantitative relationship of $N$, $N_c$, $N_{s1}$ and $N_{s0}$. They can guarantee the correct results when the correct key set is applied, and corrupt outputs when incorrect key set is inputted.
\begin{equation} \label{eq: N}
N = N_{s0} + T
\end{equation}
\begin{equation} \label{eq: Ns1}
N_{s1} = N_c+N_{s0}
\end{equation}
The netlist reveals C-SAR output $M_{out}$ and circuit output $C_{out}$ are at the different gate levels When the attacker examines the camouflaged layout. This thwarts the attacker to determine the function of C-SAR output $M_{out}$ from the temporal distribution of inputs. 

$G2$ and $G3$ are AND gates with $N_{c}+1$ inputs. When the correct key set is applied to the block, $M_{out}$ is assigned by signal $X$. $D4$ will take 2 clock cycles to output the equivalent relation of the correct key set and key inputs $K_1^{in}$ to $K_n^{in}$. Then equivalent result is passed to the camouflaged DFF array from $CD_1$ to $CD_{Nc}$. It is explained in Section \ref{sec:C_DFF} that the camouflaged DFF performs as a JTL which means the signal values from $v_1$ to $v_{N_c}$ are all identical. As a result, the output of $G2$ equals to $v_0$, the value from the $D5$. Then this signal passes $N_{s0}$ standard DFFs and the MUX to reach $M_{out}$. It takes $N_{s0}+T$ clock cycles in total to generate $M_{out}$, which has the same clock cycles as circuit output $C_{out}$ from equation (\ref{eq: N}) to guarantee the correct functions of circuit.

However, if the incorrect key set is applied, the control signal forces $M_{out}$ equal to $\Bar{Y}$. If we do not consider $G3$ first. C-SAR block takes $N_{s1}+T$ clock cycles to output $M_{out}$. In other words, path-balancing DFF array at $\Bar{Y}$ can postpone $M_{out}$ for $N_c$ clocks based on equation (\ref{eq: Ns1}), when the $C_{out}$ reaches to the XOR gate. In each attack iteration: if incorrect key set is applied at clock $t_{i}$, the $M_{out}$ can flip the circuit output $C_{out}$ at clock $t_{i+N_{s0}+N_{c}+T}$. However, incorrect key set at $t_{i+N_{c}}$ is ruled out by the SAT attack solver due to execution time in functional netlist is $T+N_{s0}$ clock cycles. If key input is correct key at clock $t_{i+N_{c}}$, it can not be eliminated as $X$ takes $T+N_{s0}$ clock cycles to $M_{out}$ which will overwrite $\Bar{Y}$ generated at $t_{i+N_{s0}+N_{c}+T}$ by incorrect key inputs at clock $t_{i}$. Hence, there are $N_c$ incorrect key sets left when all the SAT attack iterations are finished at clock $t=2^{N_{key}}$. SAT attack needs extra input patterns in number of $N_c$ to capture the rest delayed $M_{out}$. The attack process has been extended from $2^{N_{key}}$ to $2^{N_{key}}+N_c$ clock cycles to discriminate the whole incorrect key sets compared with RSAT.

Then $G3$ is implemented based on the above temporal obfuscation input design to further improve the resilience of the SAT-based attack. $G3$ causes the condition for excluding key sets to be more stringent. It scans the output value $u_1, u_2,\dots, u_{N_c}$ from $D6$ to $D{6+N_c}$ in each clock cycle. If an incorrect key set is applied to input at clock $t_i$, SAT attack must keep setting identical incorrect key set and input pattern to inputs in the next $N_{c}$ clock cycles to eliminate this incorrect key set. The process of eliminating an error key set climbs from 1 clock cycle to $N_c+1$ clock cycles. 

\subsubsection{Provably secure obfuscation analysis of C-SAR}\label{sec:secure analysis}
The resilience of the C-SAR is analyzed from the complexity of SAT-based attacks to recover the correct key. Equation (\ref{eq:RSAT}) summarizes the Boolean function of RSAT:
\begin{equation}\label{eq:RSAT}
    M_{out}(t) = F(I_i(t),K_i^{in}(t))
\end{equation}
where $I_i(t)$ and $K_i^{in}(t)$ are the primary input and key set inputs of netlists at clock cycle $t$, respectively. The output is a single bit $M_{out}(t)$. The Boolean function of C-SAR can be represented as: 
\begin{equation}\label{eq:C-SAR}
\begin{split}
    & M_i(t)= F(I_i(t),K_i^{in}(t))\\
    & M_{out}(t) = M_i(t)\cap M_i(t-1)\cdots \cap M_i(t-N_{c}+1)
\end{split}
\end{equation}
where $N_{c}$ denotes the number of camouflaged DFFs. In a SAT algorithm (Section \ref{sec:SAT}), a SAT-based attack is described by the equation (\ref{eq:SAT}), where $K^{c}$ is the correct key set:
\begin{equation}\label{eq:SAT}
F(I_i(t),K_i^{in}(t)) \oplus F(I_i(t),K^{c})
\end{equation}
The SAT-based attack selects different $I^{in}$ from the pattern set $\mathcal{I}$ in each iteration $(I^{in} \in \mathcal{I})$, and each $I^{in}$ is unique in the $\mathcal{I}$ \cite{evaluating encryption}. It indicates that the SAT-based attack only inputs each $I^{in}$ to the circuit for 1 clock cycle. However, the C-SAR only exposes incorrect key set when the $I^{in}$ and identical incorrect key set hold for $N_c+1$ clock cycles according to equation (\ref{eq:C-SAR}). Therefore, the C-SAR can immunize against SAT-based attacks. For $N_{key}$ key bits, if the SAT-based attack removes unique limitation in $\mathcal{I}$, then the complexity of attack will increase to:
\begin{gather}
\prod_{i=1}^{N_{key}} \frac{(N_{key})^{N_c+1}}{N_{key}} =\frac{(N_{key})^{N_c+1}}{N_{key}}\times\frac{(N_{key}-1)^{N_c+1}}{N_{key}-1}\dots \frac{1^{N_c+1}}{1}\notag
\end{gather}
If we consider the worst case scenario of C-SAR: the attacker modifies the SAT-based attack to a special version which extends the hold cycle of $I^{in}$ to $N_c+1$ clock cycles from 1 in each attempt of attack. The extraction of $K^{c}$ is also a discrete logarithm (DL) problem increasing exponentially with $N_{key}$. It requires $(N_c+1)*2^{N_{key}}+N_c$ clock cycles to recover $K^{c}$. We name this S-SAT attack.

\begin{table}[b]
    \centering
    \caption{
    $N_{clk}$ for 10 rounds SAT attacks \cite{evaluating encryption} to break Logic Locking, SARLock and C-SAR, when $N_{key}=3$. $NL$ is each technique integrated netlist; $R$ is the round number; "x" means SAT attack fail; C-SAR2 is C-SAR netlist attacked by S-SAT attack}
    \label{tab:avg_simulation}
    \begin{tabular}{|l|c|c|c|c|c|c|c|c|c|c|c|c|c|}
    \hline
    \diagbox{$NL$}{$R$} &1 & 2 & 3 & 4 & 5 & 6 & 7 & 8 & 9 & 10 & $avg$\\\hline
    LL      & 3 & 2 & 4 & 2 & 3 & 3 & 2 & 3 & 4 & 2 & 2.8\\\hline
    SARLock     & 8 & 8 & 8 & 8 & 8 & 8 & 8 & 8 & 8 & 8 & 8\\\hline
    C-SAR    & x & x & x & x & x & x & x & x & x & x & x\\\hline
    C-SAR2  & 17 & 17 & 17 & 17 & 17 & 17 & 17 & 17 & 17 & 17 & 17\\\hline
    \end{tabular}
\end{table}

\begin{table*}[t]
    \centering
    \caption{$N_{clk}$ for SAT attack \cite{evaluating encryption} to break Logic Locking, SARLock and C-SAR for different $N_{key}$}
    \label{tab:diff k}
    \begin{tabular}{|c||c|c|c|c|c||c|c|c|c|c||c|c|c|c|c|}
    \hline
    {} & \multicolumn {5}{c||}{LL}
    & \multicolumn {5}{c||}{SARLock}
    & \multicolumn {5}{c|}{C-SAR ($N_c$=1)}\\\hline
    \diagbox{Benchmark}{$N_{key}$} &5 &6 &7 &8 &9  &5 &6 &7 &8 &9         &5 &6 &7 &8 &9\\\hline
    74283     &6 &8 &7 &7 &8  &32 &64 &128 &256 &512 &65 &129 &257 &513 &1025\\\hline
    74182     &4 &7 &9 &8 &8  &32 &64 &128 &256 &512 &65 &129 &257 &513 &1025\\\hline
    74181     &5 &8 &7 &9 &9  &32 &64 &128 &256 &512 &65 &129 &257 &513 &1025\\\hline
    C499      &4 &6 &9 &8 &10 &32 &64 &128 &256 &512 &65 &129 &257 &513 &1025\\\hline
    C880      &7 &7 &8 &8 &9  &32 &64 &128 &256 &512 &65 &129 &257 &513 &1025\\\hline
    \end{tabular}
\end{table*}

\section{experiment results}
\subsection{Benchmark and experiment setup}
The proposed C-SAR technique is analyzed on ISCAS’85 and 74X-Series combinational benchmarks. The C-SAR superconducting circuits are implemented by JSIM (Josephson simulator), its programming format is similar to SPICE for building netlists \cite{SFQ_tool}. The JSIM generates $.raw$ files based on the netlists and simulation parameters. The Matlab tool can extract the data from these files to output signal waveform. 

In the experiments, C-SAR (Figure \ref{fig:C-RSAT}) is compared against two baselines: (1) XOR/XNOR key gates logic locking as the example in Figure \ref{fig:circuit set}.b, named LL for short. The reason to select LL as a baseline is its best resistance to the other attacks on logic locking \cite{evaluating encryption}; (2) SARLock, a SAT attack resistant logic locking technique which requires the attack iterations grow exponentially with key bits. 

For SAT attacks, a random input pattern sequence is generated to simulate the behavior of the SAT attacks to ICs. This sequence includes all possible input patterns, and each input pattern can only appear once. Since the RSFQ netlists are pipelined, the execution time depends on the gate-level $N_{gl}$ and the number of applied input patterns $N_{IP}$. If an input pattern $I^{in}$ holds $2$ clock cycles at the input, it indicates the attacker sets $I^{in}$ to circuit at the first clock, then sets $I^{in}$ again in next clock. Therefore, $N_{IP}$ equals to $2$. $N_{gl}$ is a constant dependent on benchmark and is redundant in the execution time comparison across the benchmarks. Thus, we will not include $N_{gl}$ into the $N_{clk}$ in experiment results. The large $N_{clk}$ demonstrates this technique has a high resistance to the SAT-based attacks. In section \ref{sec:result}, the area overhead and the $N_{clk}$ are reported as metrics to evaluate the performance of netlists. 
 \begin{figure}[t]
    \centering
        \includegraphics[width=0.45\textwidth]{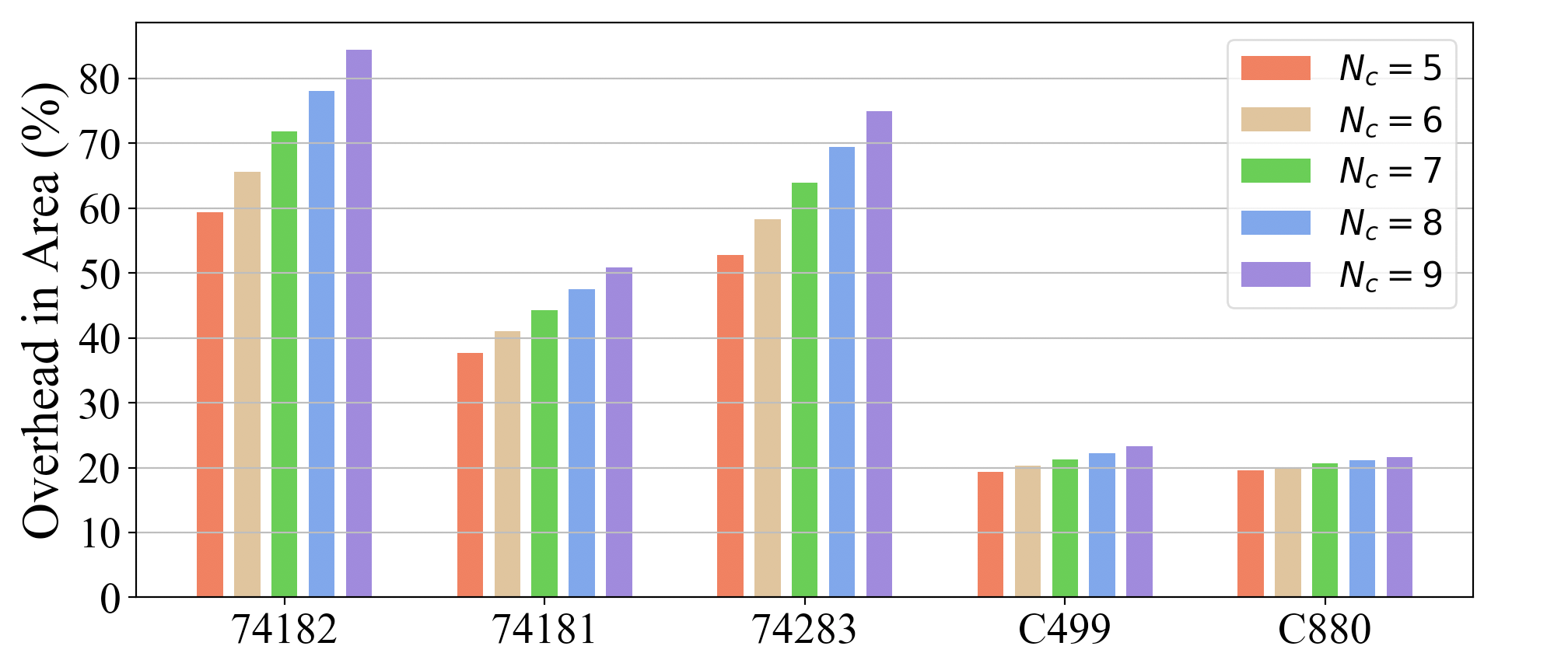}
        \caption{Area overhead for C-SAR in different key bits}
        \label{fig:overhead}
\end{figure}
\subsection{Efficiency and resilience evaluations of C-SAR}\label{sec:result}
Different input pattern sequences are applied to each technique, along with netlists, to simulate ten rounds SAT attacks in benchmark C17. It evaluates the average resilience of each approach. The number of camouflaged DFF $N_c$ in C-SAR is set to 1. Table \ref{tab:avg_simulation} reports $N_{clk}$ when key bits $N_{key}$ is $3$. SAT-based attack can easily dig out the correct key set of LL with several clock cycles, while facing SARLock, it shows stronger resilience when the attack eliminates its wrong key sets. For C-SAR, SAT-based attack could not prune any incorrect key sets as the result in Table \ref{tab:avg_simulation}, row C-SAR. Then C-SAR is attacked by S-SAT attack (explained in Section \ref{sec:secure analysis}), it can also have a significant improvement in thwarting this specially designed SAT attacks as shown in Table \ref{tab:avg_simulation}, row C-SAR2. $N_{clk}$ in C-SAR2 reaches the double of the SARLock. In the following experiments, we adopt S-SAT attack to C-SAR for evaluating its resilience.

To further evaluate the encryption strength of C-SAR, we increase the size of key bits $N_{key}$ range from 5 to 9 and the number of camouflaged DFFs $N_{c}$ from 1 to 5. In Table \ref{tab:diff k} and \ref{tab:different N_c}, the $N_{clk}$ in SARLock doubles with each increment in key bits, across all the benchmarks. However, the cost of $N_{clk}$ in C-SAR increases exponentially based on $N_{key}$ first, then has a linear relationship with $N_c+1$, finally needs another $N_{c}$ clocks to catch the rest wrong key sets. It matches the analysis in section \ref{sec:secure analysis}. These results indicate the vulnerability of LL \cite{EPIC} to SAT attacks, relative strong resilience of SARLock, and strong immunity of C-SAR in resisting SAT attacks. 

Figure \ref{fig:overhead} shows that the area overhead of C-SAR increases linearly with the size of key bits, as the number of gates inserted by C-SAR grows linearly. However, the improvement of the encryption strength of C-SAR is over exponential trend.

\begin{table}[t]
    \centering
    \caption{$N_{clk}$ for SAT attack \cite{evaluating encryption} to break C-SAR for different $N_c$, when $N_{key}=3$}
    \label{tab:different N_c}
    \begin{tabular}{|c|c|c|c|c|c|c|}
    \hline
    {} & SARLock &\multicolumn{5}{|c|}{C-SAR in different $N_c$}\\
    \hline
    Benchmark &- &1 &2 &3 &4 &5\\\hline
    C499      &8  &17 & 26 & 35 & 44 & 53\\\hline
    74181     &8  &17 & 26 & 35 & 44 & 53\\\hline
    74182     &8  &17 & 26 & 35 & 44 & 53\\\hline
    \end{tabular}
\end{table}

\section{related work}
\textbf{SARLock: SAT attack resilient logic locking:} SARLock is a SAT attack resistant logic locking technique which causes the attack to grow exponentially with the key size \cite{SARLock}. It highlights these advantages in C-SAR: Firstly, the area overheads in both approaches are similar. However, C-SAR can immunize against SAT attacks. SARLock technique can only increase the iterations of SAT attacks. Secondly, SARLock has only one fixed configuration. C-SAR can be configured differently by tuning the camouflaged DFFs.

\noindent
\textbf{Anti-SAT: mitigating SAT attack on logic locking:} Anti-SAT is another netlist technique to mitigate SAT attacks on logic locking \cite{Anti-SAT}. It provides two portions of keys: one is in the original circuit to obfuscate functionality, another portion of keys is connected to the Anti-SAT block to thwart a SAT attack \cite{Anti-SAT}. Although Anti-SAT block can be configured differently by adjusting the functionality of the logic block, the resilience performance to SAT attack is at the same level as SARLock. 

\section{conclusion}
We propose a RSFQ logic based logic locking technique, C-SAR, for immunizing against SAT-based attacks \cite{evaluating encryption}. Even under the worst case of C-SAR: facing S-SAT attacks, the results in section \ref{sec:result} show that C-SAR can improve the resilience of netlists exponentially with the key size first, then linearly with the length of camouflaged DFF array. C-SAR utilizes the DFF array to extends the required clock cycles of the attack inputs and reduces the number of key values filtered in each iteration of the attack to resist SAT-based attacks, whereas exhibits an area overhead that increases only linearly with key size. Camouflaged DFFs can also obfuscate the temporal distribution of the C-SAR inputs to resist reverse engineering attacks.

\vspace{12pt}

\end{document}